\documentclass[prl,twocolumn,superscriptaddress,nofootinbib]{revtex4-1}
\usepackage{amsfonts}
\usepackage{graphicx}
\usepackage{bm}
\usepackage{amssymb}
\usepackage{color}
\usepackage{amsmath}
\usepackage{amstext}
\usepackage{fancyvrb}
\usepackage{latexsym}
\usepackage[usenames,dvipsnames]{xcolor}
\usepackage[colorlinks=true,citecolor=Cerulean,linkcolor=RubineRed,urlcolor=Cerulean]{hyperref}

\newcommand{\beq}{\begin{equation}}
\newcommand{\eeq}{\end{equation}}
\newcommand{\beqa}{\begin{eqnarray}}
\newcommand{\eeqa}{\end{eqnarray}}

\begin{document}

\title{Probing the Full Distribution of Many-Body Observables by Single-Qubit Interferometry}
\author{Zhenyu Xu}
\affiliation{School of Physical Science and Technology, Soochow University, Suzhou 215006, China}
\affiliation{Department of Physics, University of Massachusetts, Boston, Massachusetts 02125, USA}
\author{Adolfo del Campo}
\affiliation{Donostia International Physics Center, E-20018 San Sebasti\'an, Spain}
\affiliation{IKERBASQUE, Basque Foundation for Science, E-48013 Bilbao, Spain}
\affiliation{Department of Physics, University of Massachusetts, Boston, Massachusetts 02125, USA}
\affiliation{Theoretical Division, Los Alamos National Laboratory, Los Alamos, New Mexico 87545, USA}

\begin{abstract}
We present an experimental scheme to measure the full distribution of
many-body observables in spin systems, both in and out of equilibrium, using
an auxiliary qubit as a probe. We focus on the determination of the
magnetization and the kink-number statistics at thermal equilibrium. The
corresponding characteristic functions are related to the
analytically-continued partition function. Thus, both distributions can be
directly extracted from experimental measurements of the coherence of a
probe qubit that is coupled to an Ising-type bath, as reported in [X. Peng
\textit{et al}., Phys. Rev. Lett. \textbf{114}, 010601 (2015)] for the
detection of Lee-Yang zeroes.
\end{abstract}

\maketitle

Across a continuous phase transition, a system evolves from a high-symmetry
phase to a broken symmetry phase characterized by the emergence of a new
macroscopic order. The latter can be detected by the order parameter, which
vanishes in the high-symmetry phase and acquires a non-zero value below the
critical point. A paradigmatic example is the transition between a
paramagnet and a ferromagnet in spin systems, in which the magnetization is
the order parameter.

The characterization of the order parameter statistics beyond its mean value
is motivated both by the possibility to probe its fluctuations and the quest
for the fundamental physics they may unveil, both in and out of equilibrium.
Pioneering studies in spin systems \cite%
{Binder1981,Bruce81,Bruce85,Nicolaides88} focused on the distribution of the
magnetization. The latter has applications in a wide variety of contexts
ranging from the characterization of density fluctuations in a liquid-gas
critical point \cite{BruceWilding92} to the study of turbulent fluids \cite%
{Bramwell98,AjiGoldenfeld01}. The distribution of other many-body
observables in spin systems has also proved useful. A prominent example is
the number distribution of topological defects, that is relevant to memory
devices and magnetic data storage \cite{Shinjo2013,Denisov05} and the study
of universal critical dynamics beyond the paradigmatic Kibble-Zurek
mechanism \cite{dCZ14,delcampo18}. Measuring the full distribution of
many-body observables in the laboratory is however a challenging task.

An auxiliary system, such as a single qubit, can be used as a meter in this
context.
In particular, single-qubit interferometry has been used to
characterize quantum fluids \cite{Fischer2003,Recati05,Hangleiter15,ElliottJohnson16},
determine Loschmidt echoes \cite{Quan06,Zhang08,Goold11,Knap12}, monitor
critical dynamics of decoherence \cite{Damski11}, measure work statistics
\cite{Dorner13,Mazzola13,Roncaglia14,Serra14}, the temperature of a sample
\cite{Correa15}, out-of-time order correlators \cite{Swingle16,Laura17}, and
the distribution of Lee-Yang zeroes \cite{Wei12,Wei14,Peng15}, to name some
relevant examples.

In this work, we propose an experimental protocol to measure the full
distribution of a wide class of many-body observables in a spin system
making use of a single auxiliary probe qubit. In particular, we focus on the
determination of the magnetization and kink-number distributions. To this
end, we first show that the corresponding characteristic functions at
equilibrium are given by the analytic continuation of the partition function
in classical systems. Exploiting this connection, we show that the
probability distributions can be experimentally measured (e.g. in a NMR
setting \cite{Lu2016}) by monitoring the quantum coherence of the probe
qubit that is coupled to the spin bath, that has already been demonstrated
in the laboratory \cite{Wei12,Wei14,Peng15}.


\textit{Statistics of many-body observables.---} Consider a system of $N$
interacting spins subject to a magnetic field $h$ and described by the
Hamiltonian
\begin{equation}
H_{\mathrm{s}}(\mathcal{J},h)=-\sum_{l=1}^{N}\sum_{n_{1}<\cdots
<n_{l}}^{N}\!\!\!\!J_{n_{1}\cdots n_{l}}\sigma _{n_{1}}\cdots \sigma
_{n_{l}}-h\sum_{n=1}^{N}\sigma _{n},  \label{H_Ising_g}
\end{equation}%
where the spin $\sigma $ takes values $\pm 1$ and the first term describes
arbitrary interactions among spins, including onsite disorder. The
equilibrium properties of such system can be extracted with knowledge of the
partition function in the canonical ensemble $Z(\mathcal{J},h,\beta
)=\sum_{\{\sigma =\pm 1\}}e^{-\beta H_{\mathrm{s}}(\mathcal{J},h)}$, where $%
\mathcal{J}$ denotes the set of coupling constants \{$J_{n_{1}\cdots n_{l}}$%
\}, and $\beta =1/(k_{B}T)$. A prominent instance is the Ising chain, where
the spin-spin interactions are pair-wise and restricted to nearest
neighbors, i.e., $J_{nm}=J\delta _{n,n+1} $ \cite{Goldenfeld92,Plischke06}.
The latter exhibits a phase transition between a paramagnetic and a
ferromagnetic phase, where the new macroscopic order in the broken symmetry
phase is detected by the magnetization, that is the order parameter,%
\begin{equation}
M=\sum_{n=1}^{N}\sigma _{n},  \label{M}
\end{equation}%
with possible integer values $m\in \lbrack -N,N]$. Another example concerns
the characterization of domains separated by topological defects, e.g.,
kinks. The kink number is given by%
\begin{equation}
K=\frac{1}{2}\left( N-\sum_{n=1}^{N}\sigma _{n}\sigma _{n+1}\right) ,
\label{K}
\end{equation}%
with possible integer values $k\in \lbrack 0,N]$. Without loss of
generality, in what follows we consider a many-body observable of the form%
\begin{equation}
X=a+b\sum_{\{n_{1},\cdots ,n_{l}\}}^{N}\sigma _{n_{1}}\cdots \sigma _{n_{l}},
\label{X}
\end{equation}%
which includes the magnetization $M$ (with $a=0$, $b=1$, $l=1 $) as well as
the kink number $K$ with $a=N/2$, $b=-1/2,$ $l=2$ ($\{n_{1},n_{2}\}=\{n,n+1%
\} $). In this paper, we aim at reconstructing the probability distribution $%
P(x)$ of $X$ in systems described by Hamiltonian Eq. (\ref{H_Ising_g}), this
is,
\begin{equation}
P(x)=\left\langle \delta (X-x)\right\rangle .
\end{equation}%
Here, the average denoted by $\left\langle \cdot \right\rangle
=\sum_{\{\sigma =\pm 1\}}\rho \cdot $, is taken with respect to the system
probability distribution $\rho $. The Kronecker delta $\delta (x)$ is used,
assuming that for a given spin-configuration the statistical quantity $X$
takes integer values $\{x\}$ (in the continuous case a Dirac delta function
should be used instead). By using its integral representation, the
distribution of $P(x)$ can be expressed as the Fourier transform
\begin{equation}
P(x)=\frac{1}{2\pi }\int_{0}^{2\pi }d\theta F(\theta )e^{-ix\theta }
\end{equation}%
of the characteristic function
\begin{equation}
F(\theta )=\langle e^{i\theta X}\rangle .
\end{equation}%
At equilibrium, the average is taken with respect to the canonical
distribution $\rho _{\mathrm{th}}=e^{-\beta H_{\mathrm{s}}(\mathcal{J},h)}/Z(%
\mathcal{J},h,\beta )$. The characteristic function $F(\theta )$ of $P(x)$
is then given by the analytic continuation of the partition function
\begin{equation}
F(\theta )=e^{i\theta a}\frac{Z(\mathcal{\tilde{J}},\tilde{h},\beta )}{Z(%
\mathcal{J},h,\beta )},  \label{CF}
\end{equation}%
where $Z(\mathcal{\tilde{J}},\tilde{h},\beta )=\sum_{\{\sigma =\pm
1\}}e^{-\beta H_{\mathrm{s}}(\mathcal{J},h)}e^{i\theta (X-a)}$. Here we have
introduced the effective modified set of complex-valued coupling constants $%
\mathcal{\tilde{J}}$, and the complex magnetic field $\tilde{h}$, whose
explicit forms are related to the specific statistical quantity to be
measured, as we detail below.

We note that the analytic continuation of the partition function has proved
useful in a variety of contexts including the study of phase transitions
\cite{YangLee52,LeeYang52} and the measurement of Lee-Yang zeros \cite%
{Wei12,Wei14,Peng15}, as well as the characterization of quantum chaotic
systems in relation to information scrambling \cite%
{Cotler2017,Dyer2017,delcampo17} and work statistics \cite{Chenu2017e},
among other examples.

\begin{figure}[t]
\begin{center}
\includegraphics[width=0.65\linewidth]{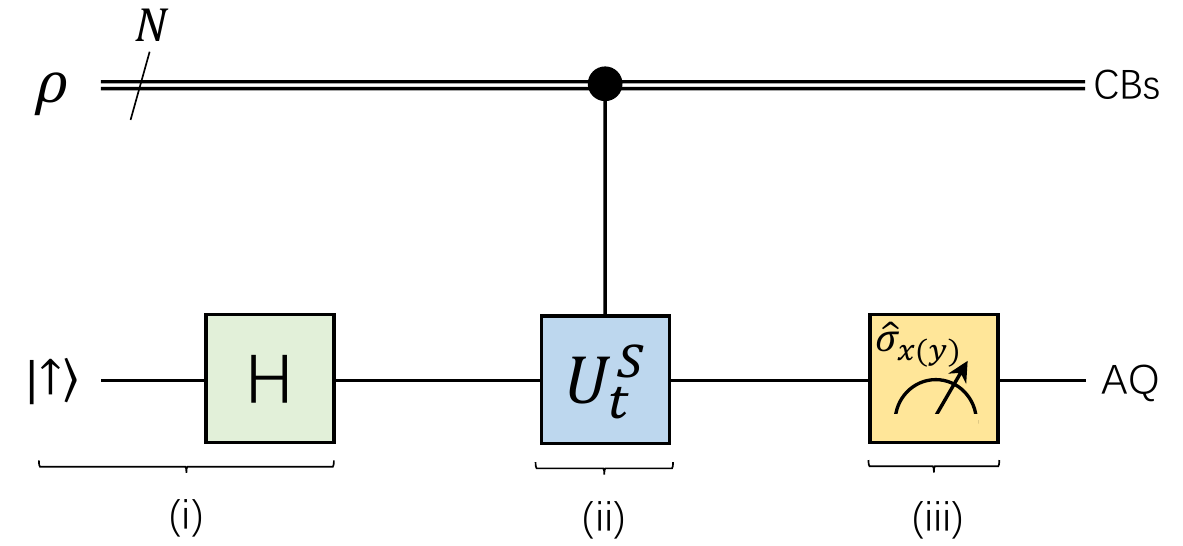}
\end{center}
\caption{\textbf{Scheme for probing the characteristic function of
statistical quantity X by a quantum simulator. }(i)\textbf{\ }Initial state
preparations. The classical spin (bit) systems (CBs) and the auxiliary qubit
(AQ) are prepared respectively in probability distribution $\protect\rho$
(or a thermal equilibrium state $\protect\rho _{\mathrm{th}}$ discussed in
the main text), and a superposition state $\left\vert +\right\rangle
=(\left\vert \downarrow \right\rangle +\left\vert \uparrow \right\rangle )/%
\protect\sqrt{2}$ by a Hadamard gate. (ii) The information of the
characteristic function is encoded into the auxiliary qubit by a simulated
unitary operator $U_{t}^{S}$ (see e.g., Fig. 2(a) and Fig. 3(a) respectively
for the detection of magnetization order parameter and kink number). (iii)
The real and imaginary parts of the characteristic function are detected by $%
\left\langle \hat{\protect\sigma} _{x}\right\rangle $ and $\left\langle \hat{%
\protect\sigma}_{y}\right\rangle $ respectively. }
\label{Scheme}
\end{figure}

\textit{Probing the characteristic function with an auxiliary qubit.---} A
schematic diagram for the detection of characteristic functions with an
auxiliary qubit is shown in Fig. \ref{Scheme}. The proposed experimental
protocol consists of the following steps:

\textit{Step 1}: 
The probe qubit is first prepared in the quantum superposition state $%
\left\vert +\right\rangle =(\left\vert \downarrow \right\rangle +\left\vert
\uparrow \right\rangle )/\sqrt{2}$ by the action of a Hadamard gate $\mathrm{%
H}=(\hat{\sigma}_{x}+\hat{\sigma}_{z})/\sqrt{2}$ ($\hat{\sigma}_{x,y,z}$
stand for Pauli matrices). The classical spin system is prepared in the
canonical thermal equilibrium state $\rho _{\mathrm{th}}$.

\textit{Step 2}: The evolution of the composite system is simulated by $%
U_{t}^{S}=\exp (-iH_{\mathrm{int}}t)$ with $H_{\mathrm{int}}=\epsilon \hat{%
\sigma}_{z}X$ ($\epsilon $ denotes a coupling constant), which can be
further written in an explicit form as
\begin{equation}
U_{t}^{S}=e^{-ita\epsilon \hat{\sigma}_{z}}\prod_{\{n_{1},\cdots
,n_{l}\}}^{N}\exp \left( -itb\epsilon \hat{\sigma}_{z}\sigma _{n_{1}}\cdots
\sigma _{n_{l}}\right) .  \label{Uts}
\end{equation}
This unitary operator can be implemented via digital quantum
simulation \cite{Lloyd96}. Then, the time-evolved state reads
\begin{equation}
\rho _{t}=\frac{1}{2}(e^{i\Omega t}\left\vert \downarrow \right\rangle
+\left\vert \uparrow \right\rangle )(e^{-i\Omega t}\left\langle \downarrow
\right\vert +\left\langle \uparrow \right\vert )\rho _{\mathrm{th}},
\end{equation}%
where $\Omega =2\epsilon X$.

\textit{Step 3}: 
The coherence of the probe spin measured by monitoring the operator $2\hat{%
\sigma}_{+}=\hat{\sigma}_{x}+i\hat{\sigma}_{y}$, i.e.,
\begin{equation}
\left\langle \hat{\sigma}_{x}\right\rangle +i\left\langle \hat{\sigma}%
_{y}\right\rangle =\sum_{\{\sigma _{n}=\pm 1\}}e^{i\mathrm{\Omega }t}\rho _{%
\mathrm{th}}=e^{i2\epsilon at}\frac{Z(\mathcal{\tilde{J}}^{\prime },\tilde{h}%
^{\prime },\beta )}{Z(\mathcal{J},h,\beta )},  \label{CFprobe}
\end{equation}%
where $Z(\mathcal{\tilde{J}}^{\prime },\tilde{h}^{\prime },\beta
)=\sum_{\{\sigma =\pm 1\}}e^{-\beta H_{\mathrm{s}}(\mathcal{J}%
,h)}e^{i2\epsilon t(X-a)}$. Parameters $\mathcal{\tilde{J}}^{\prime }$ and $%
\tilde{h}^{\prime }$ represent the effectively complex spin coupling
constants and the complex magnetic field, respectively, which are dependent
on the specific choice of $X$. If we select $2\epsilon t=\theta $, we have $%
\mathcal{\tilde{J}}^{\prime }=\mathcal{\tilde{J}}$ and $\tilde{h}^{\prime }=%
\tilde{h}$. Then Eq. (\ref{CFprobe}) is exactly the same as the
characteristic function in Eq. (\ref{CF}). After performing the Fourier
transformation, the full distribution of $X$ can be immediately
reconstructed.

In what follows we illustrate the above general scheme by analyzing two
instances of $X$: the magnetization $M$ and kink number $K$ for a
finite-temperature Ising chain.


\textit{Example 1: Measuring the full distribution of the magnetization.---}
Let us consider the Ising chain with nearest-neighbor interactions described
by the Hamiltonian
\begin{equation}
H_{\mathrm{s}}(J,h)=-J\sum_{n=1}^{N}\sigma _{n}\sigma
_{n+1}-h\sum_{n=1}^{N}\sigma _{n},
\end{equation}%
with the periodic boundary condition $\sigma _{N+1}=\sigma _{1}$. The
analytically-continued partition function can be derived by the method of
transfer matrix \cite{Plischke06}, which yields
\begin{equation}
Z(J,h,\beta )=\lambda _{-}^{N}+\lambda _{+}^{N},  \label{PF}
\end{equation}%
where $\lambda _{\pm }=e^{\beta J}\cosh (\beta h)\pm e^{-\beta J}\sqrt{%
1+e^{4\beta J}\sinh ^{2}(\beta h)}$. The magnetization, Eq. (\ref{M}), is
the order parameter. To determine its full distribution, we note that the
numerator of the characteristic function Eq. (\ref{CF}) can be written as $%
Z(J,\tilde{h},\beta )$, where $\tilde{h}=h+i\theta /\beta $ is the complex
magnetic field.

\begin{figure}[t]
\begin{center}
\includegraphics[width=0.99\linewidth]{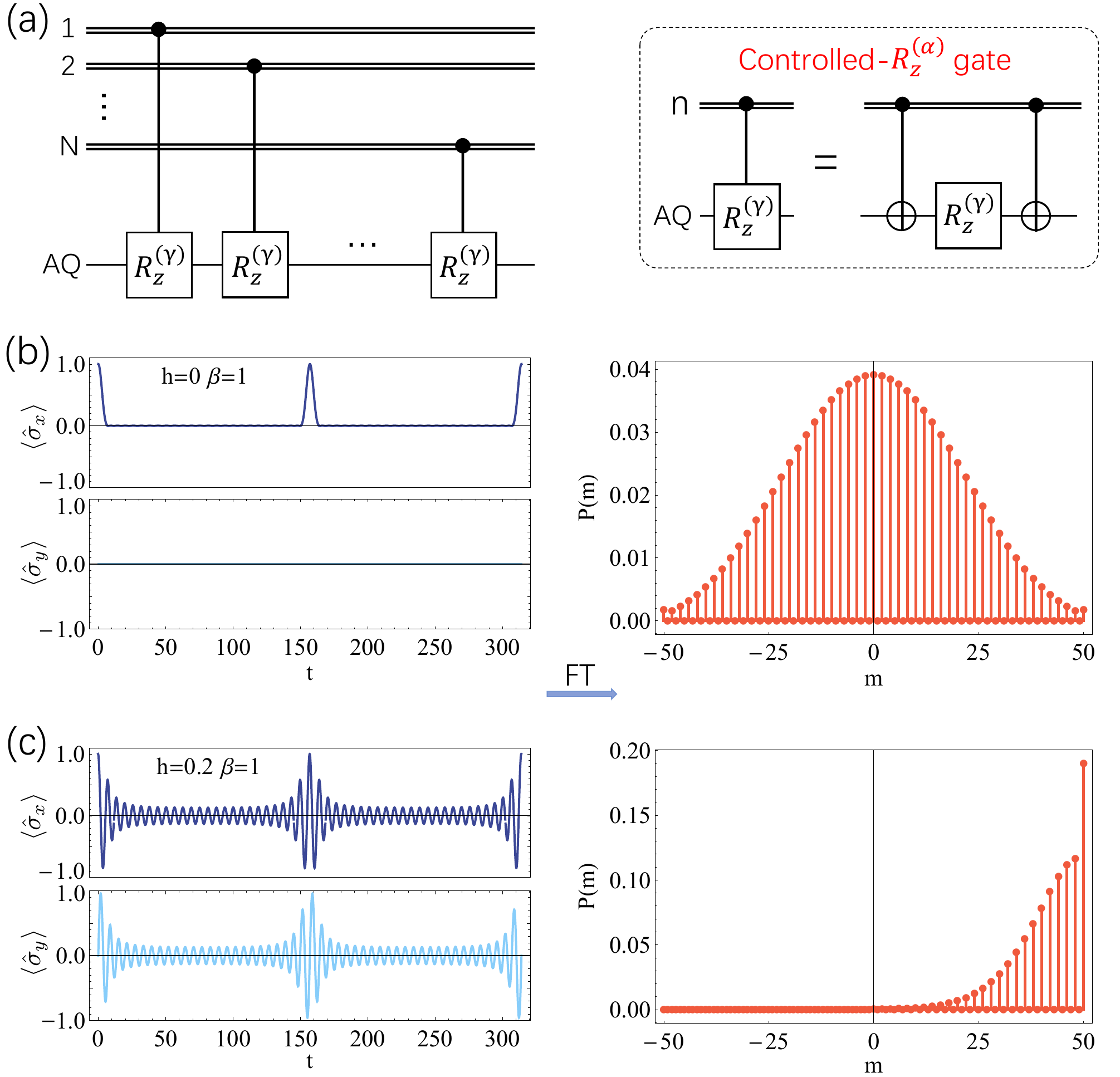}
\end{center}
\caption{\textbf{Probing the distribution of the magnetization in a nearest-neighbor Ising chain.} (a) A schematic quantum
circuit for simulating the evolution of the composite system $%
U_{t}^{S}=\prod_{n=1}^{N}\exp \left( -it\protect\epsilon \hat{\protect\sigma}%
_{z}\protect\sigma _{n}\right) $ [Eq. (\protect\ref{Ut M})]. The kernel of $%
U_{t}^{S}$ is realized by a controlled-$R_{z}^{(\protect\gamma )}$ gate ($%
\protect\gamma =2\protect\epsilon t$), with $R_{z}^{(\protect\alpha )}=\exp
(-\frac{i\protect\alpha }{2}\hat{\protect\sigma} _{z})$. Note that for the
controlled-NOT gate, the target qubit is flipped if the control classical
bit is set to $\protect\sigma_n=-1$. The characteristic function of the
magnetization distribution is detected by the coherence function of an
auxiliary qubit. For an Ising chain of $N=50$ spins at inverse temperature $%
\protect\beta =1$ with magnetic field (b) $h=0$ and (c) $h=0.2$, the real ($%
\left\langle \hat{\protect\sigma} _{x}\right\rangle $) and imaginary ($%
\left\langle \hat{\protect\sigma} _{y}\right\rangle $) parts of the
coherence function are displayed as a function of time $t\in \lbrack 0,%
\protect\pi /\protect\epsilon ]$, with the spin-qubit coupling $\protect%
\epsilon =0.01$. The right panels of (b) and (c) correspond to the
probability distribution of the magnetization obtained by Fourier transform
(FT) of the characteristic function.}
\label{Fig2_ExpProbDist}
\end{figure}

To measure the characteristic function, we use the general scheme introduced
above with the unitary
\begin{equation}
U_{t}^{S}=\prod_{n=1}^{N}\exp \left( -it\epsilon \hat{\sigma}_{z}\sigma
_{n}\right) ,  \label{Ut M}
\end{equation}%
which can be simulated with quantum logical circuit depicted in Fig. \ref%
{Fig2_ExpProbDist}(a). After measuring the real and imaginary parts of the
coherence function of the auxiliary qubit, and performing the Fourier
transformation, the distribution $P(m)$ is obtained. See examples with $h=0$
and $h=0.2$ in Fig. \ref{Fig2_ExpProbDist}(b) and (c), respectively. Note
that the magnetization varies in jumps of two units. For an even number of
spins $N$, $P(m)=0$ for odd $m$, while for odd $N$, $P(m)=0$ for even $m$.
We notice that our approach can be readily applied to long-range spin
systems as shown in the Supplemental Material \cite{SM}.

The cumulant generating function of the magnetization, which is the
logarithm of the characteristic function, admits the following expansion
\begin{equation}
\log F(\theta )=\sum_{j=1}^{\infty }\kappa _{j}\frac{(i\theta )^{j}}{j!}.
\label{cumexp}
\end{equation}%
In the large $N$ limit, one can invoke the approximation $Z(J,\widetilde{h}%
,\beta )\approx \lambda _{+}^{N}(\theta )$. Explicit computation yields the
following expressions for the first few cumulants (related to the mean,
variance, and skewness)
\begin{eqnarray}
\kappa _{1} &=&\langle M\rangle =Nu^{-1}e^{2\beta J}\sinh (\beta h), \\
\kappa _{2} &=&\mathrm{Var}(M)=Nu^{-3}e^{2\beta J}\cosh (\beta h), \\
\kappa _{3} &=&\mathrm{Skew}(M)\kappa _{2}^{3/2}=Nu^{-5}ve^{2\beta J}\sinh
(\beta h),
\end{eqnarray}%
the first two being well known (see e.g. \cite{Plischke06}), where $u=\sqrt{%
1+e^{4\beta J}\sinh ^{2}(\beta h)}$ and $v=1-e^{4\beta J}[2+\cosh (2\beta h)]
$ for short. The finiteness of $\kappa _{q}$ with $q>2$ makes $P(m)$
manifestly non-normal. In the large $N$ limit, all cumulants of $P(m)$ scale
linearly with the system size $N$. As a result, the ratio between the
amplitude of the fluctuations quantified by the root mean square $\Delta
M=\kappa _{2}^{1/2}$ and the average magnetization per spin $\langle
M\rangle $ is proportional to $\Delta M/\langle M\rangle \propto N^{-1/2}$
and fluctuations are suppressed in the thermodynamic limit. Keeping $\kappa
_{1}$ and $\kappa _{2}$ and ignoring higher-order cumulants, the probability
distribution can then be approximated by a Gaussian $P(m)=C\exp [-(m-\langle
M\rangle )^{2}/(2\mathrm{Var}(M))]$, with support on $M\in \lbrack -N,N]$,
with $C$ a normalization constant. Deviations from this limit are manifested
for nonzero values of $h$. 

\textit{Example 2: Measuring the full kink-number distribution.---} In spin
systems, kinks are localized at the interface between adjacent domains. We
next consider the distribution of the number of kinks. Considering Eq. (\ref%
{K}), the corresponding complex parameters in the analytically-continued
partition function in the numerator of Eq. (\ref{CF}) are given by $\tilde{J}%
=J-\frac{i\theta }{2\beta }$, and $\tilde{h}=h$.

%
\begin{figure}[t]
\begin{center}
\includegraphics[width=0.99\linewidth]{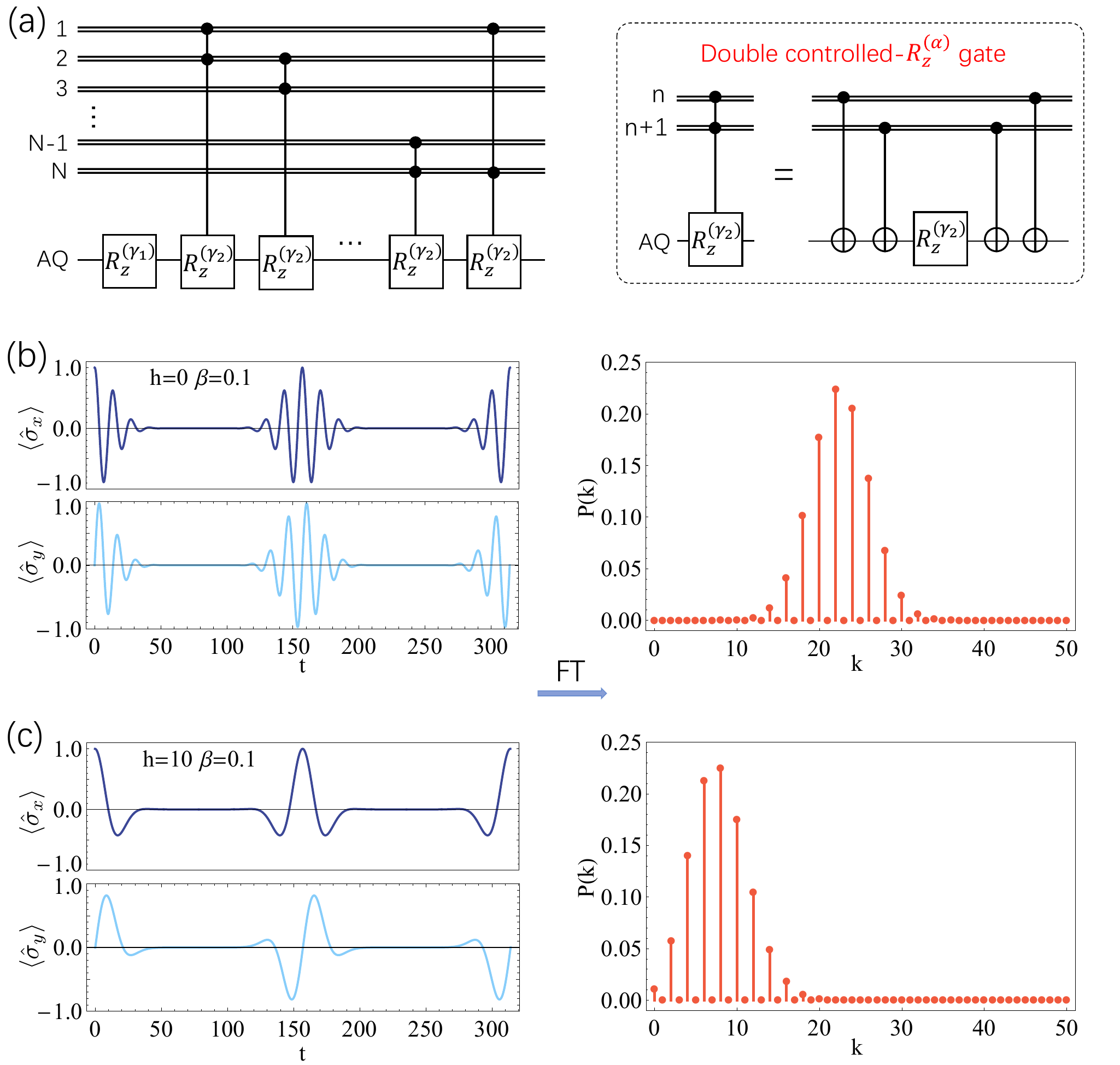}
\end{center}
\caption{\textbf{Probing the distribution of the kink number in a
nearest-neighbor Ising chain.} (a) A schematic quantum circuit for
simulating the evolution of the composite system $U_{t}^{S}=e^{-it\frac{N}{2}%
\protect\epsilon \hat{\protect\sigma} _{z}}\prod_{n=1}^{N}\exp \left( it%
\frac{\protect\epsilon }{2}\hat{\protect\sigma} _{z}\protect\sigma _{n}%
\protect\sigma _{n+1}\right) $ [Eq. (\protect\ref{Ut K})], with $\protect%
\gamma _{1}=N\protect\epsilon t$ and $\protect\gamma _{2}=-\protect\epsilon %
t $. The real ($\left\langle \hat{\protect\sigma} _{x}\right\rangle $) and
imaginary ($\left\langle \hat{\protect\sigma} _{y}\right\rangle $) parts of
the coherence function are displayed as a function of time $t\in \lbrack 0,%
\protect\pi /\protect\epsilon ]$, with the spin-qubit coupling $\protect%
\epsilon =0.01$, $N=50$, $\protect\beta =0.1$ and magnetic field $h=0$ in
(b) and $h=10$ in (c). The right parts of (b) and (c) are the probability
distribution of the kink number.}
\label{Fig3_ExpProbDist}
\end{figure}

The general scheme can be used to experimentally determine the
characteristic function, choosing
\begin{equation}
U_{t}^{S}=e^{-it\frac{N}{2}\epsilon \hat{\sigma}_{z}}\prod_{n=1}^{N}\exp
\left( it\frac{\epsilon }{2}\hat{\sigma}_{z}\sigma _{n}\sigma _{n+1}\right) ,
\label{Ut K}
\end{equation}%
which can be simulated with quantum logical circuit shown in Fig. \ref%
{Fig3_ExpProbDist}(a). For different values of $h$ and $\beta $, monitoring
the coherence of the qubit allows one to reconstruct the full kink-number
distribution $P(k)$.

The logarithm of the characteristic function of the kink-number distribution
is the corresponding cumulant generating function. For the nearest-neighbor
Ising model, approximating the partition function in the large $N$ limit by
the largest eigenvalue $Z(\tilde{J},h,\beta )\approx \lambda _{+}^{N}(\theta
)$, one finds the first cumulant, that equals the mean number of kinks
\begin{equation}
\left\langle K\right\rangle =\frac{N}{u\lambda _{+}e^{\beta J}}.
\label{kink k1}
\end{equation}%
At zero field, $h=0$, this expression reduces to $\langle K\rangle
=N/(1+e^{2\beta J})$, which varies from $0$ to $N/2$ as the temperature is
increased. The explicit expression for the variance is given by%
\begin{equation}
\mathrm{Var}(K)=\frac{N\left[ \cosh (\beta h)+2e^{3\beta J}\sinh ^{2}(\beta
h)\lambda _{+}\right] }{u^{3}\lambda _{+}^{2}},
\end{equation}%
which reduces to $Ne^{2\beta J}(1+e^{2\beta J})^{-2}$ when $h=0$. While
higher cumulants can be found making use of the expansion (\ref{cumexp})
with the corresponding characteristic function, their expression becomes
increasingly cumbersome (see e.g. the third cumulant $\kappa _{3}$ in \cite%
{SM}). At zero magnetic field, the standard deviation $\Delta K=\sqrt{%
\mathrm{Var}(K)}$ over the mean vanishes as one approaches the thermodynamic
limit with increasing system size $N$ as
\begin{equation}
\frac{\Delta K}{\langle K\rangle }=\frac{e^{\beta J}}{\sqrt{N}},
\end{equation}%
and $P(k)$ approaches the normal distribution.

Non-normal features of $P(k)$ arising from non-vanishing cumulants $\kappa
_{q}$ with $q>2$ become apparent at finite values of the magnetic field, see
Figs. \ref{Fig3_ExpProbDist}(b)-(c), as well as low temperatures.

\textit{Discussions}.\textit{--- }The scheme for probing the full
distributions of magnetization and kink number in quantum systems is similar
to the classical case. Assume the auxiliary qubit and the quantum spin
system are initially prepared in state $\left\vert +\right\rangle
\left\langle +\right\vert \otimes \rho $, where $\rho $ is an arbitrary
quantum state of the system, e.g., in or out of equilibrium. A Hadamard gate
is applied to the auxiliary qubit after performing a controlled gate $%
U_{\theta }^{S}=\left\vert \uparrow \right\rangle \left\langle \uparrow
\right\vert \otimes \exp (i\theta X)+\left\vert \downarrow \right\rangle
\left\langle \downarrow \right\vert \otimes \mathbb{I}$ on the whole system.
The state of the auxiliary qubit is then given by \textrm{tr}$_{spins}[(%
\mathrm{H}\otimes \mathbb{I)}U_{\theta }^{S}(\left\vert +\right\rangle
\left\langle +\right\vert \otimes \rho )U_{\theta }^{S\dag }(\mathrm{H}%
\otimes \mathbb{I)}]=[\mathbb{I+}\hat{\sigma}_{z}\text{Re}F(\theta )+\hat{%
\sigma}_{y}\text{Im}F(\theta )]/2$.
Thus, the real and imaginary parts of the characteristic function $F(\theta )$ can be recovered by measuring the
operators $\hat{\sigma}_{z}$ and $\hat{\sigma}_{y}$, respectively, on the
auxiliary qubit. See \cite{SM} for an alternative scheme. While the
experimental protocol can thus be adapted for quantum systems \cite{SM}, the equilibrium
relation between $F(\theta )$ and the analytically-continued partition
function is generally lost as the system Hamiltonian $H_s$ and $X$ do not
necessarily commute. Similarly, for nonequilibrium states, whether quantum
or classical, the measurement protocol applies but the connection with the
partition function is lost.

\textit{Summary.---} We have presented a general scheme to experimentally
measure the full distribution of the many-body observables in classical and
quantum systems, using an auxiliary qubit as a probe. We have demonstrated
our scheme by considering the distribution of the magnetization and the
number of kinks in classical spin systems. In this setting, the
characteristic functions of the corresponding equilibrium distributions have
been shown to be directly given by the analytic continuation of the
partition function. This connection is readily applicable to other spin
systems where the partition function in the presence of a magnetic field is
at reach, such as Heisenberg spin chains \cite{Fisher64,Cregg08}. We note
that the analytic continuation of the partition function has already been
measured experimentally in the determination of Lee-Yang zeroes in a NMR
setting \cite{Peng15}. Our proposal is within reach of
current technology. While we have focused on the determination of the
magnetization and kink-number distributions at equilibrium, we emphasize
that our scheme can be also applied to scenarios away from equilibrium. Our
findings should therefore find broad applications in the characterization of
many-body spin systems across different disciplines, including statistical
mechanics, condensed matter or magnetometry.

\textit{Acknowledgment.-} It is a pleasure to acknowledge discussions with
Fernando Javier G\'{o}mez-Ruiz, Luis Pedro Garc\'{\i}a-Pintos, Kohei
Kawabata, Geza Toth and Masahito Ueda. Funding support from the John
Templeton Foundation, UMass Boston (project P20150000029279), and the
National Natural Science Foundation of China (Grant No. 11674238) is further
acknowledged.

\bibliography{references_cl1DIM}

\pagebreak
\widetext
\begin{center}
\vspace{1cm}
\textbf{{\large Supplemental Material}}
\end{center}

\setcounter{equation}{0} \setcounter{figure}{0} \setcounter{table}{0}
\makeatletter
\renewcommand{\theequation}{S\arabic{equation}} \renewcommand{\thefigure}{SM%
\arabic{figure}} \renewcommand{\bibnumfmt}[1]{[#1]} \renewcommand{%
\citenumfont}[1]{#1}

\tableofcontents

\section{I. Higher cumulants for the distribution of kink number in the Ising
chain}

As noted in the main text, the characteristic function $F(\theta )$ of the
kink distribution $P(k)$ is given by the analytic continuation of the
partition function
\begin{equation}
F(\theta )=Z(\tilde{J},h,\beta )=\lambda _{-}^{N}+\lambda _{+}^{N},
\end{equation}%
with $\tilde{J}=J-\frac{i\theta }{2\beta }$. Using a cumulant expansion, one
can readily find an arbitrary cumulant. The exact first cumulant ($h=0$)
reads
\begin{equation}
\langle K\rangle =\frac{N}{2}e^{-\beta J}\frac{\cosh ^{N-1}(\beta J)-\sinh
^{N-1}(\beta J)}{\cosh ^{N}(\beta J)+\sinh ^{N}(\beta J)}.
\end{equation}%
The exact expressions of higher order cumulants are somewhat cumbersome.
Using the simplified expression for the partition function $Z(\tilde{J}%
,h,\beta )\approx \lambda _{+}^{N}(\theta )$, the cumulants admit simpler
expressions (see e.g. $\kappa _{1}$ and $\kappa _{2}$ in the main text). In
particular, the third cumulant, which is proportional to the skewness, reads

\begin{equation}
\kappa_3=\mathrm{Skew}(K)\kappa_2^{3/2}=\frac{Ne^{-\beta J}\left[ 5e^{2\beta
J}-(2+w)e^{2\beta J}\cosh (2\beta h)-2uw\cosh (\beta h)+4(u^{2}-1)\lambda
_{-}e^{\beta J}\cosh (\beta h)\right] }{2u^{5}\lambda _{+}^{3}},
\end{equation}%
where $w=1-8e^{8\beta J}\sinh ^{4}(\beta h)$.

One readily finds that the cumulant generating function is proportional to
the system size, as $\log F(\theta )\approx N\log \lambda _{+}^{N}(\theta )$%
, where $\lambda _{+}^{N}(\theta )$ is independent of $N$. Thus, all
cumulants of the distribution scale linearly with $N$.

\section{II. Probing magnetization order parameter and kink number in long-range
Ising chain}

The quantum circuit for the simulation of $U_{t}^{S}$ is only dependent on
the statistical quantity we are going to measure, but independent of the
specific spin system under study, i.e., it is not specific of the
short-range Ising chain. As such, it can be applied to the characterization
of the distribution of many-body observables in other experimentally
relevant spin systems. Let us consider the long-range Ising model, with the
following Hamiltonian%
\begin{equation}
H_{\mathrm{s}}\left( J,h\right) =-J\sum_{m<n}^{N}\sigma _{m}\sigma
_{n}-h\sum_{n=1}^{N}\sigma _{n},
\end{equation}%
involving all-to-all pairwise interactions of equal strength. This model
arises naturally in an NMR setting \cite{Peng15}. The corresponding
partition function is given by%
\begin{equation}
Z(J,h,\beta )=e^{\frac{N(N-1)\beta J}{2}}e^{N\beta h}\sum_{n=0}^{N}\binom{N}{%
n}e^{-2\beta hn}e^{2\beta J(n^{2}-Nn)}.
\end{equation}
In what follows we discuss the experimental protocol to characterize both
the magnetization and kink-number distributions in this system.

\subsection{A. Probing the probabilistic distribution of magnetization order
parameter}

%
\begin{figure*}[t]
\begin{center}
\includegraphics[width=0.9\linewidth]{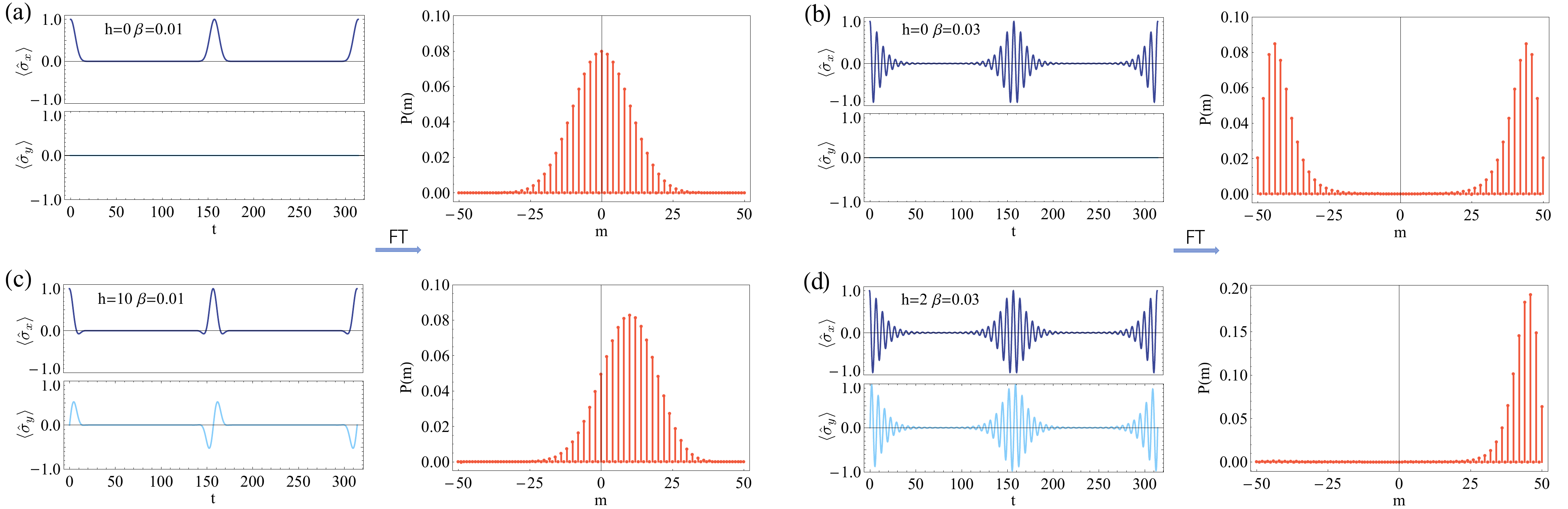}
\end{center}
\caption{\textbf{Distribution of the magnetization in a long-range Ising
model.} The real ($\left\langle \hat{\protect\sigma} _{x}\right\rangle $)
and imaginary ($\left\langle \hat{\protect\sigma} _{y}\right\rangle $) parts
of the coherence function of the auxiliary qubit are detected as a function
of time $t\in \lbrack 0,\protect\pi /\protect\epsilon ]$, with spin number $%
N=50$, the spin-qubit coupling $\protect\epsilon =0.01$, and inverse
temperature $\protect\beta =0.01$ when the magnetic field (a) $h=0$ and (c) $%
h=10$; $\protect\beta =0.03$ when the magnetic field (b) $h=0$ and (d) $h=2$%
. The probability distribution function $P(m)$ is constructed by performing
the Fourier transformation.}
\label{SM Fig M}
\end{figure*}

The characteristic function of the magnetization distribution $P(m)$ can be
written as
\begin{equation}
F(\theta )=\frac{Z(J,h+\frac{i\theta }{\beta },\beta )}{Z(J,h,\beta )},
\end{equation}%
which admits the explicit form
\begin{equation}
F(\theta )=\frac{e^{iN\theta }\sum_{n=0}^{N}e^{-2in\theta }g(n)}{%
\sum_{n=0}^{N}g(n)},
\end{equation}%
where $g(n)=\binom{N}{n}e^{-2\beta hn}e^{2\beta J(n^{2}-Nn)}$. Use of the
cumulant expansion readily yields the first few cumulants of $P(m)$%
\begin{eqnarray}
\kappa _{1}=\left\langle M\right\rangle &=&N-\frac{2G_{1}(N)}{G_{0}(N)}, \\
\kappa _{2}=\mathrm{Var}(M) &=&4\left[ \frac{G_{2}(N)}{G_{0}(N)}-\left(
\frac{G_{1}(N)}{G_{0}(N)}\right) ^{2}\right] , \\
\kappa _{3}=\mathrm{Skew}(M)\kappa _{2}^{3/2} &=&-8\left[ \frac{G_{3}(N)}{%
G_{0}(N)}-3\frac{G_{1}(N)G_{2}(N)}{G_{0}(N)^{2}}+2\left( \frac{G_{1}(N)}{%
G_{0}(N)}\right) ^{3}\right] ,
\end{eqnarray}%
where $G_{\alpha }(N)=\sum_{n=1}^{N}n^{\alpha }g(n)$.

The experimental proposal for probing the characteristic function is similar
to \textit{Example 1} in the main text, with numerical simulations shown in
Fig. (\ref{SM Fig M}).

\subsection{B. Probing the probabilistic distribution of kink number}

%
\begin{figure}[t]
\begin{center}
\includegraphics[width=0.6\linewidth]{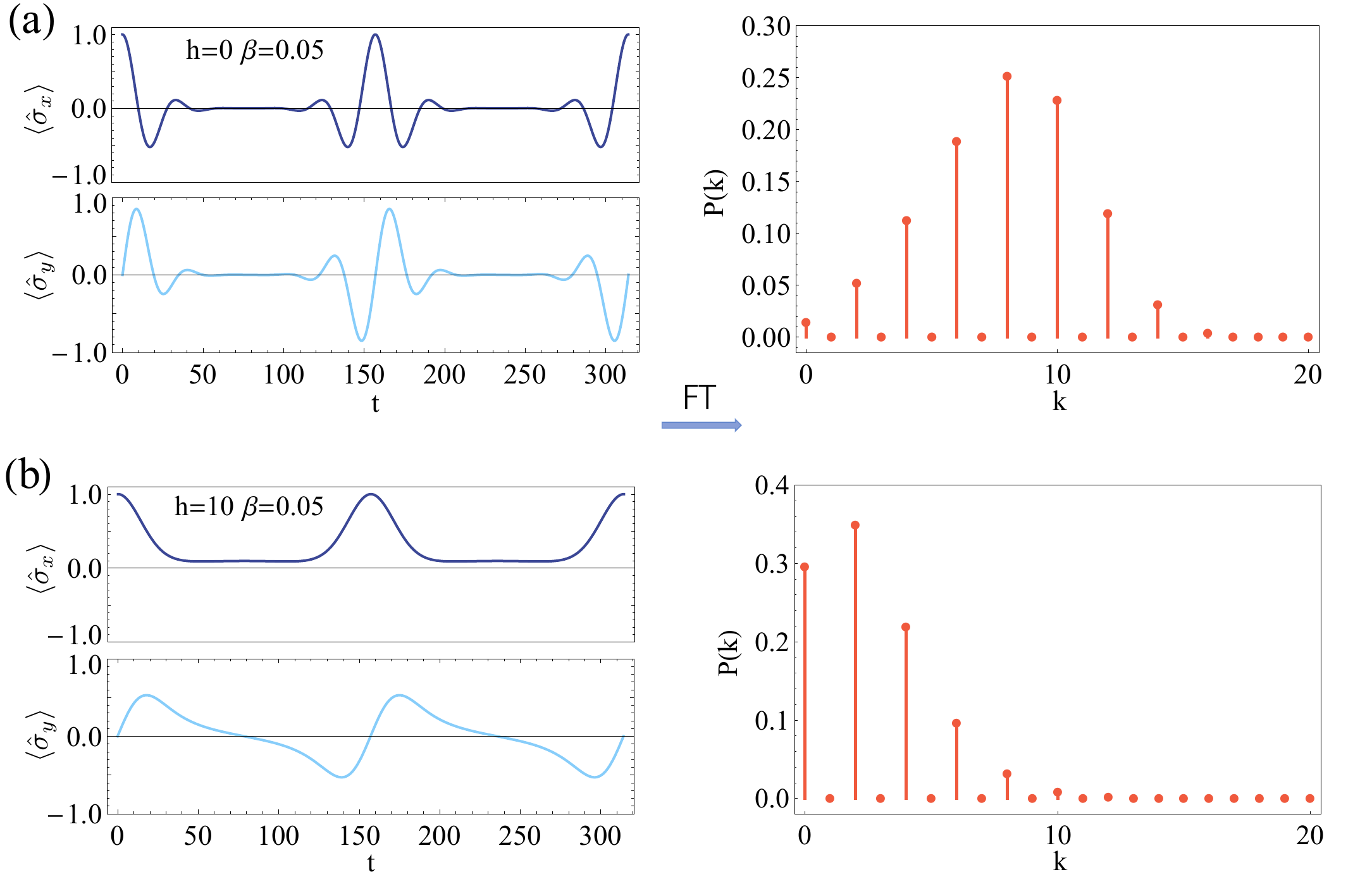}
\end{center}
\caption{\textbf{Distribution of the kink number in a long-range Ising model.%
} The real ($\left\langle \hat{\protect\sigma} _{x}\right\rangle $) and
imaginary ($\left\langle \hat{\protect\sigma} _{y}\right\rangle $) parts of
the coherence function of the auxiliary qubit are displayed as a function of
time $t\in \lbrack 0,\protect\pi /\protect\epsilon ]$, with spin number $%
N=20 $, the spin-qubit coupling $\protect\epsilon =0.01$, inverse
temperature $\protect\beta =0.05$, and the magnetic field (a) $h=0$ and (b) $%
h=10$. The probability distribution function $P(k)$ is shown after Fourier
transformation. }
\label{SM Fig K}
\end{figure}

The characteristic function is given by%
\begin{equation}
F(\theta )=e^{i\frac{N\theta }{2}}\frac{Z(\mathcal{\tilde{J}},h,\beta )}{%
Z(J,h,\beta )},
\end{equation}%
where
\begin{equation}
\mathcal{\tilde{J}}=\left\{
\begin{array}{ll}
J & n\neq m+1 \\
J-\frac{i\theta }{2\beta } & n=m+1%
\end{array}%
\right.
\end{equation}

The experimental proposal for probing the characteristic function is similar
to \textit{Example 2} in the main text, with numerical simulations shown in
Fig. (\ref{SM Fig K}).

\section{III. Probing the characteristic function for quantum spin systems}

\subsection{A. Quantum circuit}

\begin{figure}[t]
\begin{center}
\includegraphics[width=0.45\linewidth]{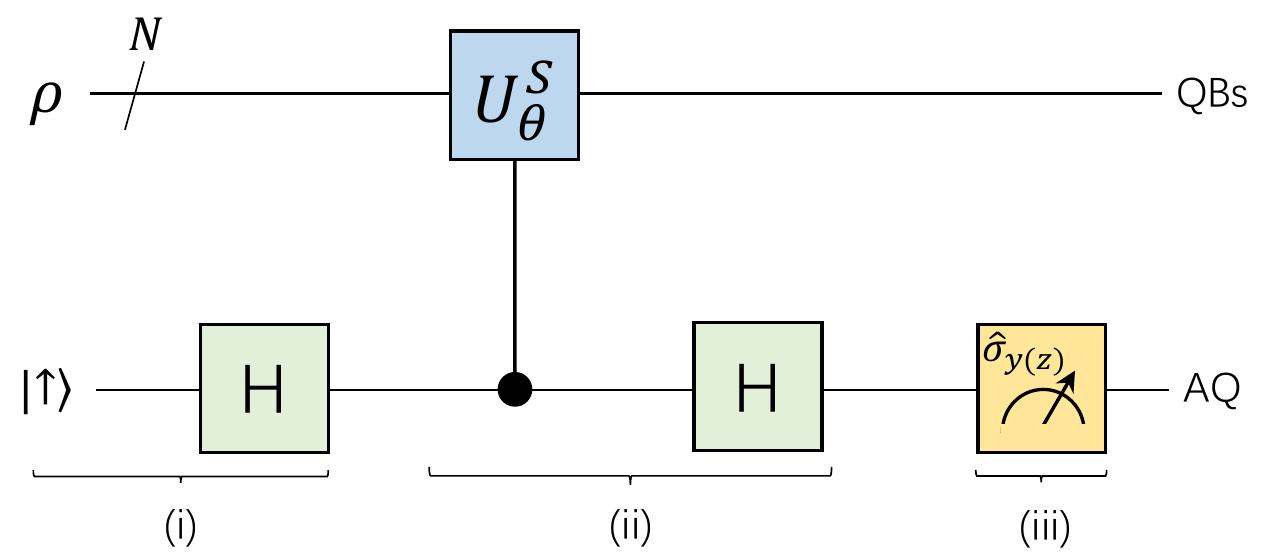}
\end{center}
\caption{\textbf{Scheme for probing the characteristic function of quantum
spin systems by a quantum simulator. }(i)\textbf{\ }Initial state
preparations. The quantum spin (qubit) systems (QBs) and the auxiliary qubit
(AQ) are prepared respectively in state $\protect\rho $, and a superposition
state $\left\vert +\right\rangle =(\left\vert \downarrow \right\rangle
+\left\vert \uparrow \right\rangle )/\protect\sqrt{2}$ by a Hadamard gate.
(ii) The information of the characteristic function is encoded into the
auxiliary qubit by a simulated unitary operator $U_{\protect\theta }^{s}$
and a Hadamard gate. (iii) The real and imaginary parts of the characteristic
function are detected by $\left\langle \hat{\protect\sigma}_{z}\right\rangle
$ and $\left\langle \hat{\protect\sigma}_{y}\right\rangle $ respectively. }
\label{SchemeQ}
\end{figure}

The scheme for probing the characteristic function for quantum spin systems
is shown in Fig. (\ref{SchemeQ}), which is slightly different from the
classical case in the main text.

\subsection{B. Probing the magnetization distribution via Loschmidt echo}

We have seen that the characteristic function of the magnetization is
related to the analytic continuation of the partition function. In quantum
systems, the later can be measured in a variety of quantum platforms
including quantum simulators of spin systems and NMR experiments. Given a
Hamiltonian $H$ with eigenvalues $E_{n}$ and energy eigenstates $|n\rangle $%
, the state of the system can be initialized in the coherent quantum
superposition of the form $|\psi (0)\rangle =\sum_{n}\sqrt{p_{n}}|n\rangle $
where $p_{n}=\exp (-\beta E_{n})/Z(\beta )$ and $Z(\beta )=\sum_{n}\exp
(-\beta E_{n})$ denotes the partition function. This superposition evolves
into $|\psi (t)\rangle =\sum_{n}\exp [-(\beta /2+it)E_{n}]|n\rangle /\sqrt{%
Z(\beta )}$, and the survival amplitude reads $\langle \psi (0)|\psi
(t)\rangle =Z(\beta +it)/Z(\beta )$. Choosing as a Hamiltonian the Ising
chain, the survival amplitude can be written as $\langle \psi (0)|\psi
(t)\rangle =Z(\beta ,h-it/\beta )/Z(\beta ,h)$, this is, in terms of the
analytically continued partition function for an Ising chain with a
complex-valued magnetic field, and can be measured experimentally via a
Loschmidt echo \cite{Peng15}. Its Fourier transform directly yields the full
distribution of the magnetization.

\section{IV. Simulation efficiency and error analysis}

In this section, we provide a brief analysis on the simulation efficiency
and possible experimental errors.

\subsection{A. An estimation of simulation efficiency}

In the main text, the evolution of the composite system is given by $%
U_{t}^{S}=\exp (-iH_{\mathrm{int}}t)$ with
\begin{equation}
H_{\mathrm{int}}=\epsilon \hat{\sigma}_{z}X=\epsilon a\hat{\sigma}%
_{z}+\epsilon b\hat{\sigma}_{z}\sum_{\{n_{1},\cdots
,n_{l}\}}^{N}H_{\{n_{1},\cdots ,n_{l}\}},
\end{equation}%
where $\epsilon $ denotes a coupling constant, and $H_{\{n_{1},\cdots
,n_{l}\}}=\sigma _{n_{1}}\cdots \sigma _{n_{l}}$ is a many-body spin
interaction. Since the spin system is classical, all $H_{\{n_{1},\cdots
,n_{l}\}}$ commute and $U_{t}^{S}$ can be directly decomposed as
\begin{equation}
U_{t}^{S}=e^{-iH_{\mathrm{int}}t}=e^{-ita\epsilon \hat{\sigma}%
_{z}}\prod_{\{n_{1},\cdots ,n_{l}\}}^{N}\exp (-itb\epsilon \hat{\sigma}%
_{z}H_{\{n_{1},\cdots ,n_{l}\}}),  \label{ut}
\end{equation}%
with no approximation, and each $e^{-itb\epsilon \hat{\sigma}%
_{z}H_{\{n_{1},\cdots ,n_{l}\}}}$ can be realized by digital simulations
with logical gates (as shown in the main text). Note that although there is
no simulation error in this case, operational errors in the lab cannot be
avoided (see an analysis of possible experimental errors in Sec. \ref{SecIVb}%
).

In the discussion in the main text, we also consider quantum spin systems.
The distribution of many-body observables
\begin{equation}
X=a+b\sum_{\{n_{1},\cdots ,n_{l}\}}^{N}H_{\{n_{1},\cdots ,n_{l}\}},\text{
with }H_{\{n_{1},\cdots ,n_{l}\}}=\hat{\sigma}_{\alpha _{1};n_{1}}\cdots
\hat{\sigma}_{\alpha _{l};n_{l}},\text{ (}\alpha _{n_{j}}=x,y,z\text{),}
\end{equation}%
can still be detected by coupling to a single probe qubit and monitoring its
coherence. The scheme requires to simulate $U_{\theta }^{S}=\left\vert
\uparrow \right\rangle \left\langle \uparrow \right\vert \otimes e^{i\theta
X}+\left\vert \downarrow \right\rangle \left\langle \downarrow \right\vert
\otimes \mathbb{I}$, in which the operator $e^{i\theta X}$ can be further
written in an explicit form as%
\begin{equation}
e^{i\theta X}=e^{e^{i\theta a}}e^{i\theta b\sum_{\{n_{1},\cdots
,n_{l}\}}^{N}H_{\{n_{1},\cdots ,n_{l}\}}}.  \label{QX}
\end{equation}%
We note that, if we only consider the magnetization $M=\sum_{n=1}^{N}\hat{%
\sigma}_{z;n}$ and the kink number $K=\frac{1}{2}(N-\sum_{n=1}^{N}\hat{\sigma%
}_{z;n}\hat{\sigma}_{z;n+1})$, there will also be no simulation error since
the items in the sum of Eq. (\ref{QX}) all commute. However, for general
many-body observables, each $H_{\{n_{1},\cdots ,n_{l}\}}$ may not commute.
To realize such scheme in digital quantum simulations, we have to perform
some approximations. One method is to employ Suzuki-Trotter expansion with $%
m $ steps. For any $\varepsilon >0$, $m$ can always be chosen sufficiently
large to ensure that $e^{i\theta X}$ is well approximated by the simulator
with an error no greater than $\varepsilon $, i.e.,

\begin{equation}
\left\Vert e^{i\theta X}-e^{e^{i\theta a}}\left( \prod_{\{n_{1},\cdots
,n_{l}\}}^{N}e^{\frac{i\theta bH_{\{n_{1},\cdots ,n_{l}\}}}{m}}\right)
^{m}\right\Vert \leq \varepsilon .  \label{simulation error}
\end{equation}%
This \textquotedblleft digital\textquotedblright\ approach to quantum
simulation [Eq. (\ref{simulation error})] aimed at probing the statistics of
general many-body observables is efficient since the number of the
combination in $\{n_{1},\cdots ,n_{l}\}$ is around $N^{n_{l}}/n_{l}!$ ($%
n_{l}\ll N$), which is a polynomial function of $N$ \cite{Lloyd96}.


\subsection{B. Experimental error analysis}

\label{SecIVb}

The analysis above shows that if the spin chain systems are classical (or
quantum spin chains with observables such as the magnetization and the kink
number), there will be no simulation errors (i.e., $U_{\mathrm{simulation}%
}\equiv U_{\mathrm{exact}}$). However the operational errors in experiment
can not be neglected, and may influence the accuracy of the final full
distribution with an error propagation
\begin{equation}
\text{Experimental errors}\rightarrow \text{Characteristic Function }F\text{(%
}\theta \text{)}\rightarrow \text{Distribution }P.
\end{equation}%
The main experimental errors we consider here stem from the imperfect
implementation of logical gates and the decoherence of the auxiliary qubit.

In NMR systems, the universal logical gates (i.e., single qubit rotation and
CNOT gates) can currently be realized by the gradient ascent pulse
engineering (GRAPE) technique with very high precision (pulses fidelity
around 99.95\%) \cite{Lu2016}. The imperfection errors can be neglected
within such a high accuracy. For the auxiliary qubit, the $T_{2}^{\ast }$
can be of the order of several seconds. A CNOT gate and single qubit
rotation gate can be implemented within several milliseconds \cite{Lu2016}.
Therefore, it is possible to implement hundreds of gates before $T_{2}^{\ast
}$ has elapsed. For example, if we adopt the trimethyl phosphite system
(consisting of $9$ $^{1}$H spins as the spin chain and one $^{31}$P nuclear
spin as the auxiliary probe qubit) in the Lee-Yang zeros experiment \cite%
{Peng15}, the total number of universal gates is $27$ [$=$($1$ single qubit
rotation gate$+2$ CNOT gates)$\times 9$] for the magnetization.
These gates can be finished within around 300 milliseconds, which is much less than the $T_{2}^{\ast }$ of the
auxiliary qubit. Therefore, our scheme is available with current NMR experimental techniques.

\begin{figure}[tbp]
\centering{}\includegraphics[width=6.5in]{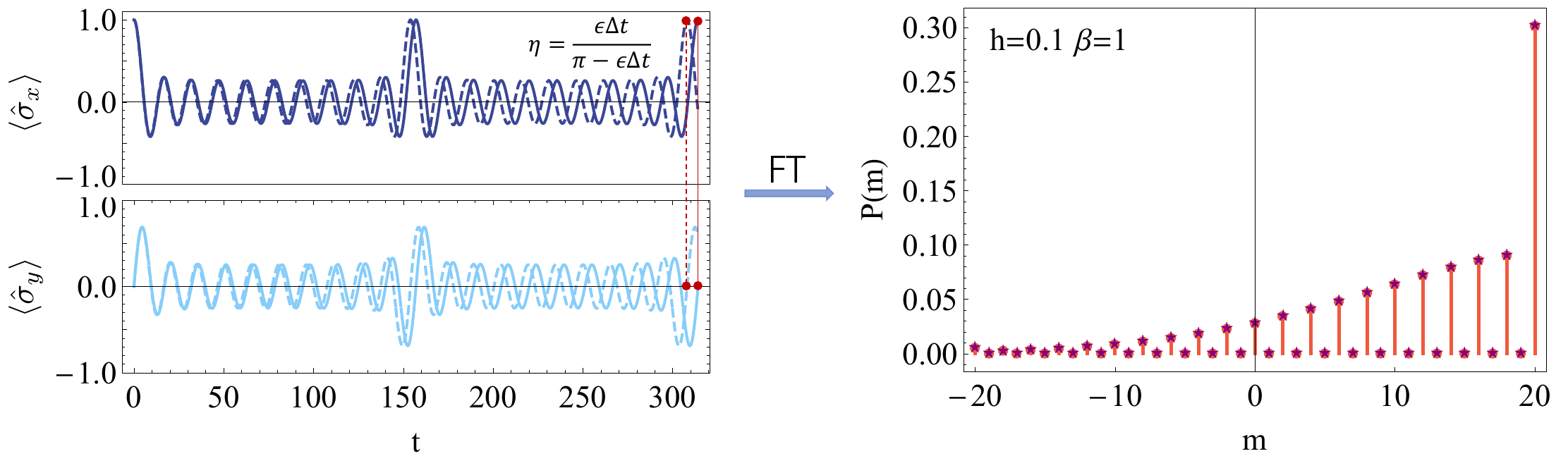}
\caption{\textbf{Error analysis for imperfect implementation of logical gates%
}. Example: Probing the distribution of the magnetization order parameter in
a nearest-neighbor Ising chain ($N=20$, $h=0.1$, $\protect\beta =1$, and $%
\protect\epsilon =0.01$). Left: the real ($\left\langle \hat{\protect\sigma}%
_{x}\right\rangle $) and imaginary ($\left\langle \hat{\protect\sigma}%
_{y}\right\rangle $) parts of the characteristic function are displayed as a
function of time $t\in \lbrack 0,\protect\pi /\protect\epsilon ]$ with no
operational errors of logical gates (solid curves) and with errors $\protect%
\eta $ (dashed curves). The error $\protect\eta $ can be evaluated by
measuring the $\Delta t$ (the time interval between the solid and dashed red
lines). Right: The probability distribution of the magnetization obtained by
Fourier transform (FT) of the characteristic function with no errors (red
dots) and by the modified FT [Eq. (\protect\ref{MFT})] with errors (purple
stars). Obviously, such kind of operational errors can be circumvented in
experiment. }
\label{SM_Fig_error}
\end{figure}


Note that even when the gate errors can not be neglected, there is a method
to circumvent such errors. Every kernel $\exp \left( -itb\epsilon \hat{\sigma%
}_{z}\sigma _{n_{1}}\cdots \sigma _{n_{l}}\right) $ in Eq. (\ref{ut}) can be
realized by a controlled rotation gate (as shown in the main text). We
assume the error of such gate is $\delta =\eta \epsilon $, then $\epsilon
^{\prime }=\epsilon +\delta =(1+\eta )\epsilon $. Following the operation
steps in the main text, the coherence of the probe spin is then given by
\begin{equation}
\left\langle \hat{\sigma}_{x}\right\rangle +i\left\langle \hat{\sigma}%
_{y}\right\rangle =e^{i2\epsilon at}\frac{Z(\mathcal{\tilde{J}}^{\prime },%
\tilde{h}^{\prime },\beta ,\epsilon ^{\prime })}{Z(\mathcal{J},h,\beta
,\epsilon )},  \label{CFprobe}
\end{equation}%
where $Z(\mathcal{\tilde{J}}^{\prime },\tilde{h}^{\prime },\beta ,\epsilon
^{\prime })=\sum_{\{\sigma =\pm 1\}}e^{-\beta H_{\mathrm{s}}(\mathcal{J}%
,h)}e^{i2\epsilon ^{\prime }t(X-a)}$. If we set $2\epsilon t=\theta $, Eq. (%
\ref{CFprobe}) is the final characteristic function (including the errors of
logical gates). In real experiment, we may directly perform the frequency
domain measurements to get the full distribution or adopt the indirect
measurement. For the latter, the distribution of X (with errors) can be
reconstructed by performing the Fourier transformation. However, if we know
the error of such logical gate, there is one method to circumvent such
errors with the following revised Fourier transform
\begin{equation}
P(x)=\frac{1+\eta }{2\pi }\int_{0}^{\frac{2\pi }{1+\eta }}d\theta F[\theta
(1+\eta )]e^{-ix\theta (1+\eta )},  \label{MFT}
\end{equation}%
with which we can reconstruct the $P(x)$. In Fig. (\ref{SM_Fig_error}), we
illustrate such scheme with the example of measuring the distribution of
magnetization ($\eta =0.02$ as an example).

\end{document}